\documentclass[%
preprint,
showpacs,
preprintnumbers,
 amsmath,amssymb,
 aps,
]{revtex4-1}

\usepackage{graphicx}
\usepackage{dcolumn}
\usepackage{bm}

\begin{document}

\preprint{}

\title{On electroweak baryogenesis in the littlest Higgs model with \textit{T} parity}

\author{Sahazada Aziz}
\email{$aziz_bu$@rediff.com}
\author{Buddhadeb Ghosh}
\email{ghoshphysics@yahoo.co.in}
\affiliation{Department of Physics, Center for Advanced Study, University of Burdwan, Burdwan-713104, India.}%
\date{\today}

\begin{abstract}
We study electroweak baryogenesis within the framework of the littlest Higgs model with \textit{T} parity. This model has shown characteristics of a strong first-order electroweak phase transition, which is conducive to baryogenesis in the early Universe. In the \textit{T} parity symmetric theory, there are two gauge sectors, viz., the \textit{T}-even and the \textit{T}-odd ones. We observe that the effect of the \textit{T}-parity symmetric interactions between the \textit{T}-odd and the  \textit{T}-even gauge bosons on gauge-higgs energy functional is quite small, so that these two sectors can be taken to be independent. The \textit{T}-even gauge bosons behave like the Standard Model gauge bosons, whereas the  \textit{T}-odd ones are instrumental in stabilizing the Higgs mass. For the \textit{T}-odd gauge bosons in the symmetric and asymmetric phases and for the \textit{T}-even gauge bosons in the asymmetric phase, we obtain, using the formalism of Arnold and McLerran, very small values of the ratio, (Baryon number violation rate/Universe expansion rate). We observe that this result, in conjunction with the scenario of inverse phase transition in the present work and the value of the ratio obtained from the lattice result of sphaleron transition rate in the symmetric phase, can provide us with a plausible baryogenesis scenario.
\end{abstract}

\pacs{98.80Cq, 12.15Ji, 12.60Cn, 12.60Fr.}
\keywords{Early Universe; Electroweak Baryogenesis; Beyond Standard Model.}

\maketitle
\section{Introduction}
The mystery of the observed baryon-antibaryon asymmetry [1, 2] of the Universe has not yet been unambiguously resolved. A study of baryogenesis usually has to incorporate Sakharov's three criteria [3] : (i) Baryon number violation, (ii) C and CP violation, (iii) Departure from thermal equilibrium. The first two conditions are needed to create the matter-antimatter asymmetry and the third one is necessary to retain that asymmetry, or, in other words, to prevent the erasure of that asymmetry. In the Standard Model (SM) of particle physics, large baryon number violation is possible at high temperature by sphaleron  transitions [4-9] between the degenerate vacua of the SU(2) gauge field. Thus, the first criterion of Sakharov is met in the SM. However, it is difficult to satisfy  the second and the third criteria in the SM as, in this model, the CP violation is too low to explain the observed baryon to entropy ratio [10] and a strong first-order electroweak phase transition (EWPT), necessary for the thermal out-of-equilibrium condition, can be obtained only for ${m_H}<32\,GeV$ [1], whereas the current experimental lower bound is ${m_H} \simeq 115 GeV$. These difficulties have been overcome in models having extended Higgs sectors, such as the Minimally Supersymmetric Standard Model (MSSM) [11-15], its extensions [16-20] and the Two-Higgs Doublet Model (THDM) [21, 22].      
      
Among the models for the new physics at the $TeV$ scale, the littlest Higgs model (${L^2}$HM) [23] and its version with T parity (LHT) [24-27] are economical and popular ones. In LHT, the aspects of CP violation have been studied [28] quite extensively in recent years. This model belongs to the class of very few BSMs having Non-Minimal Flavor Violation (NMFV) effects. New CP and flavor violation arise in this model due to heavy gauge boson mediated interaction between \textit{T}-odd mirror fermions and light SM-like fermions. The signatures of this violation have already been seen in LHCb experiments [28].

To examine the third criterion of Sakharov, finite-temperature calculations are required.  Although indications of strong first-order EWPT have been reported in a number of papers involving Supersymmetric [11-20], Two-Higgs Doublet [21, 22] and Extra-dimensional models [29-31], the needed finite-temperature calculations in Little Higgs Models are, so far, very few [32-34]. However, such studies should be quite interesting as, some of these [32, 34] have exhibited a not-so-common and intriguing feature of finite-temperature effects, viz., nonrestoration of symmetry at high temperature [32, 34-38], which might reveal new aspects of cosmological baryogenesis.
          
In this work, we examine the prospects of baryogenesis in the LHT in the light of the third criterion of Sakharov. The present work is motivated by our earlier finding [34] of features of a strong first-order EWPT in the global structure of the effective potential in the LHT, in association with a nonrestoration of symmetry at high temperature. Here, we calculate the baryon number violation rates for the \textit{T}-even and the \textit{T}-odd gauge bosons. We can find a new scenario of baryogenesis, where the \textit{T}-even particles are efficacious in producing very large amount of baryon number violation and the  \textit{T}-odd ones check the wash-out.

In section 2, we present  the gauge-higgs  sector of the LHT. In section 3, we examine the effect of the interaction between the \textit{T}-even and \textit{T}-odd gauge bosons on the sphaleron energy. In section 4 we calculate baryon number violation rates. Finally, section 5 contains discussions, whereby we elaborate the baryogenesis scenario, and then some concluding remarks.    
\section{Gauge-higgs sector in LHT}
The gauge sector in the ${L^2}$HM is contained in the Lagrangian,
\begin{eqnarray} 
L&=& \frac{{{f^2}}}{8}Tr({\partial _\mu }\Sigma  - i\sum\limits_{j = 1}^2 {[{g_j}W_{j\mu }^a(Q_j^a\Sigma
+ \Sigma Q_j^{aT})}  + {{g'}_j}{B_{j\mu }}({Y_j}\Sigma  + \Sigma {Y_j})]) 
\nonumber\\&& \times({\partial ^\mu }\Sigma  - i\sum\limits_{j = 1}^2 {[{g_j}W_j^{a\mu }(Q_j^a\Sigma 
+  \Sigma Q_j^{aT})}  + {{g'}_j}B_j^\mu ({Y_j}\Sigma  + \Sigma {Y_j})])^\dag
\end{eqnarray}
where $f$ is a scale $\sim$ 1 $TeV$ and in LHT $f$ $\sim$ .5 $TeV$. In Eq.(1) there are two SU(2) and two U(1) gauge groups instead of one each as in the Weinberg-Salam theory. However, invariance of $\cal L$ under the \textit{T} parity operations: ${W_1} \leftrightarrow {W_2}$  and ${B_1} \leftrightarrow {B_2}$, will require ${g_1} = {g_2}$, ${g'_1} = {g'_2}$ and therefore, ${g_{1,2}} = \sqrt 2 g$, ${g'_{1,2}} = \sqrt 2 g'$, where $g$ and $g'$ are SM weak and hypercharge gauge couplings. With these, after the tree-level explicit symmetry breaking at the TeV scale by the vacuum condensate ${\Sigma _0}$ [34], we shall have \textit{T}-odd heavy and \textit{T}-even light gauge bosons [27],
\begin{equation}
W_H^a = \frac{1}{{\sqrt 2 }}(W_2^a - W_1^a),{B_H} = \frac{1}{{\sqrt 2 }}({B_2} - {B_1}),W_L^a = \frac{1}{{\sqrt 2 }}(W_2^a - W_1^a),B_L = \frac{1}{{\sqrt 2 }}({B_2} + {B_1})
\end{equation}
where, $H(L)$ refers to the heavy (light) gauge bosons. ${W_L}$ and ${B_L}$ which get mass by EWSB correspond to SM gauge bosons. As in the SM, physical neutral gauge bosons are obtained via the mixing of the neutral partners of ${W_L}$ and ${B_L}$ with the Weinberg angle, ${\theta _W}$, which is given in terms of the coupling constants as, $\tan {\theta _W} = g'/g$. Spherically symmetric sphaleron will require ${\theta _W} = 0$ and therefore ${g'_{1,2}} = 0$. In that case, only the SU(2) sectors in (1) will be operative. In the present work, we shall assume that ${\theta _W} = 0$. This assumption is justified by the observation [4] that the correction to the sphaleron energy due to the ${\theta _W}$ nonzero case is quite small: it is $0.6\% $ when $\lambda  = 0$ to 0.96\% when $\lambda  = \infty $, $\lambda$  being the higgs  quartic self-coupling parameter. Following the same spirit, in the present case also, we shall consider only the SU(2) sector.

The \textit{T}-odd gauge bosons have masses both in the symmetric phase (SP) and broken phase (BP) of EWSB, because they get mass by explicit symmetry breaking while the Lagrangian is gauged. The \textit{T}-even gauge bosons  become massive after the EWSB by the Coleman-Weinberg mechanism. As a general procedure, we calculate the masses of the  \textit{T}-even and  \textit{T}-odd gauge bosons in the BP, using the vacuum condensate $\Sigma$ as in Ref.34 and get in the leading order in \textsl{s},
\begin{equation} {M_{W_H^1}} = {M_{W_H^2}} = {M_{W_H^3}} = fg\sqrt {1 - \frac{{{s^2}}}{2}} ,{M_{W_L^1}} = {M_{W_L^2}} = {M_{W_L^3}} = \frac{1}{{\sqrt 2 }}fgs
\end{equation}
where $s = sin\left( {\frac{h}{{\sqrt 2 f}}} \right)$, $h$ being the physical higgs field.
In the SP at $s=0$ we have the temperature-independent masses ${M_{W_H^1}} = {M_{W_H^2}} = {M_{W_H^3}} = fg$  and ${M_{W_L^1}} = {M_{W_L^2}} = {M_{W_L^3}} = 0$. Here the heavy gauge boson mass is the one obtained  by explicit symmetry breaking under \textit{T} parity by the vacuum condensate ${\Sigma _0}$ [34], the light gauge bosons remaining massless at all temperatures. In the BP, both the light and heavy gauge bosons have h-dependent masses given in Eq.(3). It may be noted that, if $h=v$ ( the SM VEV at zero temperature), then from (3) we find that the light gauge bosons show the SM mass, ${M_W} = \frac{1}{2}gv$ in the leading term of the sine function.

The Higgs quartic self coupling constant $\lambda$, which is an important parameter in determining the sphaleron energy and transition rate,  may be obtained numerically from the global higgs potential by the formula,
\begin{equation}
\lambda  = \frac{1}{{4!}}{\left( {\frac{{{\partial ^4}{V^{(1)}}(h)}}{{\partial {h^4}}}} \right)_{h = 1.1TeV}}
\end{equation}
where, ${V^{(1)}}(h)$ is the temperature-independent one-loop-order effective potential, which is responsible for EWSB in $L^2$HM by Coleman-Weinberg mechanism. However, $\lambda$ should be small enough for the validity of perturbative calculations and to be consistent with a not-so-large higgs mass. We find in our calculation that we can get small values of $\lambda$ by reasonable choice of the UV completion factors [23, 34]. The detailed procedure of how the UV completion factors are determined in our calculations is given in Ref.34. Note that in Eq.(4),  $h=1.1$ $TeV$  (or $s=1$) is the value of the physical higgs field where the strong first-order EWPT is observed in the global structure of the finite-temperature effective potential [34]. It may be noted that the \textit{T} parity forbids a non-zero VEV for the triplet $\phi$ field [27]. Thus, the $\phi$  field does not come in the consideration of phase transion in the present case. 
\section{A model calculation of sphaleron energy and the role of \textit{T}-odd-\textit{T}-even gauge boson interaction}
Sphaleron is the time-independent static solution of the gauge-scalar lagrangian. The benchmark of our calculation is to make use of the sphaleron transition rate formula [5] of a pure SU(2) gauge theory. In LHT there are two $SU(2)$ groups: one for \textit{T}-even light gauge boson field  and another for \textit{T}-odd heavy gauge boson field.  \textsl{T}-invariant interactions of the types ${W_L}{W_L}{W_H}{W_H}$ and ${W_L}{W_H}{W_H}$ are present between the two sectors [39]. So, before using the classic formula one has to make sure that the contribution of the interaction terms is markedly small. To verify this fact we have constructed a toy model comprising the lagrangian of a pure SU(2) theory plus a mass term for $W_H$, as it gets mass from explicit symmetry breaking of the gauge group SU(5) down to ${[SU(2) \otimes U(1)]_1} \otimes {[SU(2) \otimes U(1)]_2}$ .

The gauge-higgs energy in the model is as follows 
\begin{equation}
E = \int {( - \frac{1}{{2{g^2}}}} Tr({F_{ij}}{F_{ij}}) + {({D_i}\Phi )^\dag }({D_i}\Phi ) + \lambda {({\Phi ^\dag }\Phi  - \frac{{{v^2}}}{2})^2} + \frac{1}{2}{M_{{W_H}}}^2{W_{Hi}}{W_{Hi}}){d^3}x
\end{equation}
where, ${F_{ij}} = {\partial _i}{W_j} + {\partial _j}{W_i} + [{W_i},{W_j}]$, is the time-independent gauge boson tensor.
The first three terms in the integrand of (5) are for both the \textit{T}-even and \textit{T}-odd sectors. ${W_i}(x) =  - ig\frac{{{\sigma ^\alpha }}}{2}W_i^\alpha$ (i=1,2,3 is space index, $\alpha$= 1,2,3 is $SU(2)$ weak isospin index.
${D_i} = {\partial _i}{I_2} - {W_i}(x)$ is the covariant derivative, $\Phi$ is the scalar higgs field, ${M_{{W_H}}} \approx gf$ is the heavy gauge boson mass term.

The spherically symmteric ansatze [4] for the gauge boson and higgs scalar are respectively,
\begin{eqnarray}
{W_i}{\sigma ^\alpha } &=&  - \frac{{2i}}{g}{f_1}(gvr){\partial _i}{U^\infty }{({U^\infty })^{ - 1}},
\nonumber\\\Phi  &=& \frac{v}{{\sqrt 2 }}{f_2}(gvr){U^\infty }\left( {\begin{array}{*{20}{c}}
0\\
1
\end{array}} \right),
\end{eqnarray}
where ${f_1}$ and ${f_2}$ are funtion of the radial distance $r$, and ${U^\infty } = \frac{1}{r}\left( {\begin{array}{*{20}{c}}
z&{x + iy}\\
{ - x + iy}&z
\end{array}} \right)$. 
Using these ansatze and putting $k=gvr$, a dimensionless parameter, we can have the full energy functional as
\begin{eqnarray}
E = \int \begin{array}{l}
[\frac{{32\pi v}}{{{k^2}g}}({(2{f_1}(k)(1 - {f_1}(k)))^2} + {k^2}{(f'(k))^2}) + \frac{{8\pi v}}{g}({f_2}{(k)^2}{({f_1}(k) - 1)^2}\\
 + \frac{1}{2}{k^2}{f_2}{(k)^2}) + \frac{{\lambda v}}{{{g^3}{k^2}}}{({f_2}{(k)^2} - 1)^2} + \frac{{32\pi {f^2}}}{{gv}}{f_1}{(k)^2}]dk
\end{array}
\end{eqnarray}
Now, if we add \textsl{T} parity invariant interaction terms of the possible two types ${W_L}{W_L}{W_H}{W_H}$ and ${\partial _i}{W_L}{W_H}{W_H}$ to the hamiltonian the term which will be added to the integrand of E is $- \frac{1}{{2{g^2}}}\left( {\frac{{20{v^2}{g^2}}}{{{k^2}}}} \right){f_1}{(k)^4}$.
To evaluate the energy integral we use two ansatze [4] of ${f_1}$ and ${f_2}$ which are 
\begin{eqnarray}
{f_1}(k) &=& 1 - \frac{4}{{m + 4}}Exp[\frac{{(m - k)}}{2}],
\nonumber\\{f_2}(k) &=& 1 - \frac{n}{{\sigma n + 1}}\frac{1}{k}Exp[\sigma (n - k)], k \ge m,n,
\end{eqnarray}
where $\sigma  = \sqrt {\frac{{2\lambda }}{{{g^2}}}} $  and $m,n$ are numbers.\\ 
\begin{figure}[ht]     
\includegraphics[scale=.8]{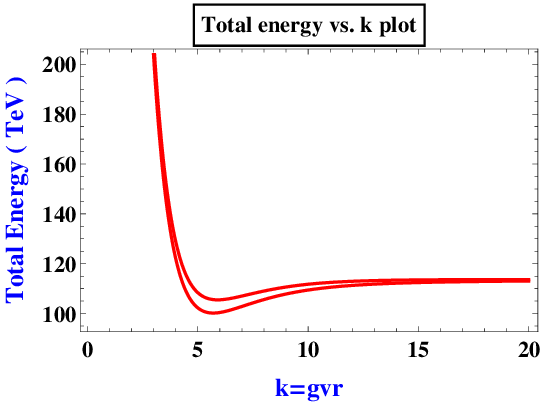}
\caption{Upper plot shows total gauge-higgs energy without interaction and lower plot shows total energy including interaction between \textit{T}-even and \textit{T}-odd gauge bosons.}
\label{fig:FIG.1}
\end{figure}
In Fig.1 we have plotted total gauge-higgs energy and have shown the effect of interaction. If we include the interactions the total energy shows only less than 1\% change, which clearly hints that the effect of the \textit{T}-parity invariant interaction has negligible effect in sphaleron energy calculations. So the two $SU(2) \otimes U(1)$ sectors can be treated independent of one another and one can make use of sphaleron transition rate formula of a pure $SU(2)$ theory reliably.     
The expression for energy functional for the ${W_H}$ is given by 
\begin{eqnarray}
E({W_H}) = \int \begin{array}{l}
[\frac{{32\pi v}}{{{k^2}g}}({(2{f_1}(k)(1 - {f_1}(k)))^2} + {k^2}{(f'(k))^2}) + \frac{{8\pi v}}{g}({f_2}{(k)^2}{({f_1}(k) - 1)^2}\\
 + \frac{1}{2}{k^2}{f_2}{(k)^2}) + \frac{{\lambda v}}{{{g^3}{k^2}}}{({f_2}{(k)^2} - 1)^2} + \frac{{32\pi {f^2}}}{{gv}}{f_1}{(k)^2}]dk .
\end{array} 
\end{eqnarray}
Simalrly the total hamiltonian or energy functional for $W_L$ is given by
\begin{eqnarray}
E({W_L}) = \int \begin{array}{l}
\frac{1}{2}gv[\frac{{32\pi }}{{{k^2}{g^2}}}({(2{f_1}(1 - {f_1}))^2} + {k^2}{({f_1}')^2}) + \frac{{8\pi }}{{{g^2}}}({f_2}^2{({f_1} - 1)^2}\\
 + \frac{1}{2}{k^2}{({f_2}')^2}) + \frac{{2\lambda }}{{{g^4}}}{k^2}{({f_2}^2 - 1)^2}]dk
\end{array} 
\end{eqnarray}
To get sphaleron energy for the ${W_L}$ and ${W_H}$ we have to minimize the energy functional with respect to $m,n$ and integrate it upto radius $r \sim \frac{1}{{{M_W}}}$ where $k \ge m,n$.
 
Taking the VEV at $v=1.1$ $TeV$ one can get the sphaleron energies to be 33.48 $TeV$ and 36.71 $TeV$ respectively for the light and heavy gauge bosons in the BP and 29.23 $TeV$ for the heavy gauge boson at SP. In this calculation the value of $\lambda$ is obtained from Eq.4 viz., $\lambda =0.8$ which is comparable with the value of ${g^2}$, where $g$ is 0.9.
\section{Thermal gauge boson masses and baryon number violation rates}
Here, we have calculated the thermal gauge boson masses using the one-loop-order finite-temperature effective potential [34]. The expression of the thermal gauge boson mass squared can be written as,
\begin{equation}
M_{{W_{H,L}}}^2(T) = M_{{W_{H,L}}}^2 + \frac{aT^4}{2{\pi ^2}{f^2}}\int\limits_0^{4\pi f/T} {x^2}log[1-Exp(- \sqrt {{x^2} + M_{{W_{H,L}}}^2/{T^2}} )]dx
\end{equation}
where,$M_{{W_H}}^2 \equiv M_{{W_H}}^2(T = 0)$ which is $\frac{1}{2}{f^2}{g^2}$ for $s = 1$ in the  BP and  ${f^2}{g^2}$ for $s = 0$ in the SP and $M_{{W_L}}^2 \equiv M_{{W_L}}^2(T = 0)$ which is $\frac{1}{2}{f^2}{g^2}$ for $s = 1$ in the  BP and 0 for $s = 0$ in SP [see Eqs.(3)].
We note here that, in the present formalism, the sphaleron transition rate and baryon number violation rate for the \textit{T}-even gauge bosons cannot be calculated in the symmetric phase, where their masses are zero. Perturbative approach fails here because of infrared divergences. The transition rate for this case has been calculated [40] in the lattice using the auxiliary field method.  
The sphaleron transition rate [5, 41] is,
 \begin{equation}
 \frac{\Gamma }{V} = 4\pi {\omega _ - }{({\tilde g^2}T)^3}{N_{tran}}{N_{rot}}\frac{{\kappa (\lambda /{g^2})}}{{{\kappa _{\max }}}}\exp [ - {E_{sph}}(T)/T]  
\end{equation}
 and the baryon number violation rate [5],
 \begin{equation}
 \frac{1}{{{N_B}}}\frac{{d{N_B}}}{{dt}} =  - \left( {\frac{{13}}{{2{\pi ^2}}}} \right){n_f}{\left[ {\frac{{{{(2{M_W(T)})}^2}}}{{{\alpha _W}}}} \right]^3}{T^{ - 6}}{\omega _ - }
 {N_{tran}}{N_{rot}}\frac{{\kappa (\lambda /{g^2})}}{{{\kappa _{\max }}}} \exp [ - {E_{sph}}(T)/T]  
\end{equation}
where, ${n_f}$ is the number of fermion family which is 4 in LHT, ${\omega _ - }$ is the rate of decay in small fluctuations around the sphaleron, which is a function of $\lambda /{\tilde g^2}$. ${N_{tran}}$ and ${N_{rot}}$ are  normalization factors related to the translational and rotational degrees of freedom of the sphaleron and $\kappa$ is the determinant associated with small fluctuations around the sphaleron. Eqs.(12) and (13) are valid in the range of temperature T lying between, ${M_W}(T)$ and ${M_W}(T)/{\alpha _W}$.Temperature dependent sphaleron energy ${E_{sph}}(T)$ is obtained using eqns. 9, 10 and 11.
  
${\omega _ - }$ in the unit of $\tilde gv$ can be obtained from the expression,
\begin{equation}
\omega _ - ^2 = [0.5143 + 0.3794(\lambda /{\tilde g^2}) - 0.0644{(\lambda /{\tilde g^2})^2}+ 0.00379{(\lambda /{\tilde g^2})^3}]{(\tilde gv)^2} 
\end{equation}
which is a fit to the corresponding plotted graph in Ref.6 in the range, $\lambda /{\tilde g^2} \approx 0.1$ to 10.

The product ${N_{tran}}{N_{rot}}$ as a function of $\lambda /{\tilde g^2}$ can be obtained from the equation [42],  
\begin{equation}
{N_{trans}}{N_{rot}} \cong 86 - 5\ln (\lambda /{\tilde g^2}) 
\end{equation}
The expression for $\kappa (\lambda /{\tilde g^2})$ can be obtained from Refs.9 and 41 as, 
\begin{equation}
\kappa  = 0.0229\exp ( - 0.13/{(\lambda /{\tilde g^2})^2} + 0.65/(\lambda /{\tilde g^2}) - 0.09(\lambda /{\tilde g^2}))
\end{equation}
with, $\kappa _{\max } = \exp ( - 3)$.

For getting the ratio of the baryon number violation rate to the Universe expansion rate (i.e., the Hubble parameter), viz., $T(d{N_B}/{N_B}dT)$, we convert temperature to time using the dynamically-generated relation [43] in the radiation dominated early Universe, $t = 2.42 \times {10^{ - 12}}g_*^{ - 1/2}\left( {\frac{1}{{{T^2}}}} \right)$ sec. 
where, T is in $TeV$ and in the $TeV$ scale ${g^*} = 145.75$ in LHT. For T= 1 $TeV$, $t \approx{10^{-13}}$ sec. 
\begin{figure}[ht]
\includegraphics[scale=.70]{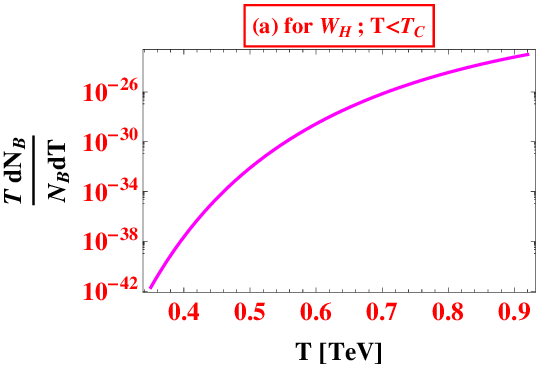}
\includegraphics[scale=.75]{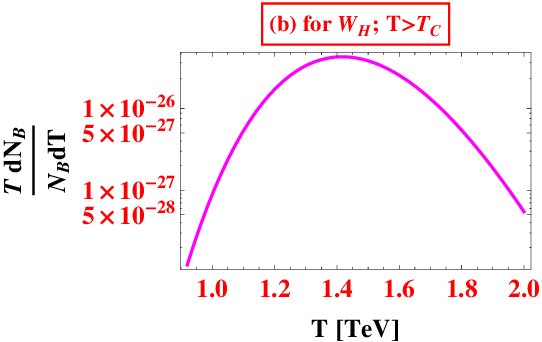}
\includegraphics[scale=.75]{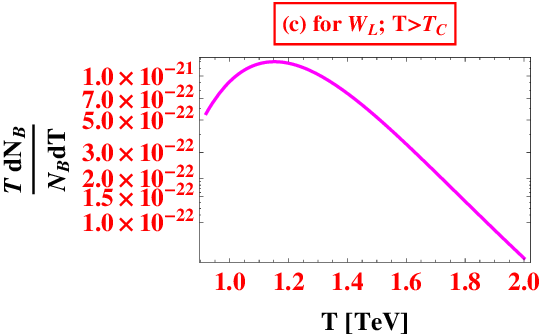}
\caption{Ratio of the baryon number violation rate to the Universe expansion rate, (a) in the symmetric phase, (b) in the asymmetric phase for the \textit{T}-odd gauge boson and (c) in the asymmetric phase for \textit{T}-even gauge boson.}
\label{fig:FIG.2}
\end{figure} 
In Fig.2 we have plotted $T(d{N_B}/{N_B}dT)$ vs. T, for the \textit{T}-odd gauge bosons for temperatures both below  and above  ${T_c}$, the transition temperature of EWPT, and that for \textit{T}-even gauge boson above ${T_c}$. It is to be noted that the value of ${T_c}$ in our case is 0.925 $TeV$ [34].The graphs in Figs 2(a) and 2(b) clearly show that around the EWPT, the contribution of the \textit{T}-odd gauge bosons in the LHT towards  the ratio of the baryon number violation rate to the expansion rate of the Universe is negligibly small. In particular, at T = 1 $TeV$ in the asymmetric phase, the baryon number violation rate is $\sim{10^{- 9}}$ $\sec^{-1}$ and the Universe expansion rate is $\sim{10^{13}}$ $\sec^{-1}$.

For the massless \textit{T}-even particles in the symmetric phase, the lattice result [40] gives, $\Gamma  \sim {10^{ - 6}}{T^4}$. From this, we get [44], $\frac{1}{{{N_B}}}\frac{{d{N_B}}}{{dt}} \approx \frac{{13}}{2}{n_f}\frac{\Gamma }{{{T^3}}} \sim {10^{ - 5}}T \sim {10^{22}}$ $\sec ^{ - 1}$ (1 $TeV \simeq 1.5 \times {10^{27}}$ $\sec^{-1}$). Thus, the baryon number violation rate is ${10^{31}}$ times higher for the \textit{T}-even particles than the \textit{T}-odd ones. 
\section{Discussion and conclusions}
We find that at the value  $h$=1.1 $TeV$ of the physical higgs field, where the phase transition is observed, both the  \textit{T}-even and \textit{T}-odd gauge bosons have the same mass (viz.,$\frac{1}{{\sqrt 2 }}fg$) in the broken phase ($T > {T_C}$) and the baryon number violation rate is seen to be negligibly small and effectively no baryon number violation takes place in the broken phase at high temperature. In the symmetric phase ($T < {T_C}$), the  \textit{T}-even gauge boson becomes massless, the \textit{T}-odd ones remaining massive with mass $fg$. In this phase,  we may consider bubbles inside which the higgs VEV is near zero. These bubbles would move in the background of asymmetric phase and the massless \textit{T}-even particles and the  massive \textit{T}-odd particles will be inside these bubbles. The sphalerons  due to the \textit{T}-odd particles  act as spectators so far as baryon number violation is concerned. Input from non-perturbative SM lattice result [40] implies that the baryon number violation rate is 31 orders of magnitude higher for the \textit{T}-even particles than that for the \textit{T}-odd ones. Since the phase transition is strongly first order the bubbles will be thin-walled and the CP-violating scattering of the \textit{T}-even particles will take place on the bubble walls due to interaction with the Higgs field. However, this situation will continue from the temperature scale $T \sim 1 TeV$ to $T \sim 0.1 TeV$ (which corresponds to the time scale: $t\sim{10^{ - 13}}$ sec to $t \sim {10^{ - 11}}$ sec), when there will be smooth crossover [45] from a massless  to a massive condition for the \textit{T}-even particle. The two-step baryogenesis process ends at this point.

It should be quite interesting to dwell on the significance of our work in the cosmological perspective. Accepting the maximum reheating temperature [43] after inflation to be much higher than $T_C$ here, the universe may have gone from a broken phase to a symmetric phase through a \textit{non-standard} EWPT at the $TeV$ scale, which is a strong first-order one, and the \textit{extreme thermal non-equilibrium situation} (giving rise to violation of time-reversal symmetry and therefore CP non-invariance)  associated with this transition has been quite efficacious in preventing the washout of the generated baryon-antibaryon asymmetry at the electroweak scale.
   
Our calculations are based on the existence of two independent $SU(2)$ gauge sectors for the \textit{T}-odd and the \textit{T}-even particles as verified by our calculation in section 3 and SM-like anomalies in the corresponding fermionic sectors. Such anomalies have not yet been investigated so far within the LHT. On the other hand, leptogenesis has been studied recenlty in ${L^2}$HM [46] and in other little Higgs models [47-49]. Thus, there is scope to obtain baryogenesis via leptogenesis in these models. We must note here that, as we are dealing with an inverted phase transition, the nature of the bubbles must be different from what they are in the SM. Specifically, the bubbles are expected to shrink and ultimately disappear at the time of the cross-over, leaving a net baryon number in the broken phase. However, the detailed dynamics of such bubbles is yet to be studied and outside the scope of the present paper. 

In conclusion, in the present work we have explored the possibility of electroweak baryogenesis in the LHT model. This is the first work on electroweak baryogenesis in a model where inverse phase transition occurs and has thus new features in it. Nonetheless, we can get a clear picture of baryogenesis in the new scenario. We find that the  \textit{T}-odd gauge bosons play crucial role in checking the washout. Discovery of these new particles is expected in the Large Hadron Collider and the future International Linear Collider experiments, for which the production cross sections have been estimated [50-52]. 
    
\section*{Acknowledgments}
The authors thank AseshKrishna Datta of Harishchandra Research Insitute, Alahabad, India for many illuminating discussions.The authors acknowledge the University Grants Commission, Government of India, for granting a major research project [F. No. 32-36/2006(R)] under which part of the present work has been done.  S.A. thanks the Council of Scientific and Industrial Research, Government of India, for granting a Senior Research Fellowship.

\end{document}